\documentstyle[12pt,psfig,draft,lscape,array]{nature-ejb}

\def\simlt{\mathrel{\hbox{\rlap{\hbox{\lower4pt\hbox{$\sim$}}}\hbox{$<$}}}}
\def\simgt{\mathrel{\hbox{\rlap{\hbox{\lower4pt\hbox{$\sim$}}}\hbox{$>$}}}}

\def\lesssim{\mathbin{\lower 3pt\hbox 
      {$\rlap{\raise 5pt\hbox{$\char'074$}}\mathchar"7218$}}} 
\def\gtrsim{\mathbin{\lower 3pt\hbox
      {$\rlap{\raise 5pt\hbox{$\char'076$}}\mathchar"7218$}}}

\newcommand{\exo}{EXO\,0748$-$676}

\newcommand{\rxte}{\textit{RXTE}}
\newcommand{\exos}{\textit{EXOSAT}}

\newcommand{\xmm}{\textit{XMM}-Newton}

\newcommand{\fedd}{\mbox{$F_{\rm Edd}$}}

\begin{document}

\title{\Large \bf \exo\ Rules out Soft Equations of State for Neutron
Star Matter}

\author{
   F.~\"Ozel\\ 
     {\footnotesize Department of Physics, University of Arizona, 1118
     E. 4th St,
      Tucson, AZ, 85704, USA}
}

\date{\today}{}
\headertitle{\exo\ Rules out Soft Equations of State for Neutron Star Matter}
\mainauthor{\"Ozel}

\summary{The interiors of neutron stars contain matter at very high
densities, in a state that differs greatly from those found in the
early universe or achieved at terrestrial experiments\cite{hands}.
Matter in these conditions can only be probed through astrophysical
observations that measure the mass and radius of neutron stars with
sufficient precision\cite{lp}. Here I report for the first time a
unique determination of the mass and radius of the neutron star \exo,
which appears to rule out all the soft equations of state of neutron
star matter. If this object is typical, then condensates and
unconfined quarks do not exist in the centers of neutron stars.}

\maketitle

The neutron star source \exo\ has repeatedly shown thermonuclear X-ray
bursts. It has also shown, first\cite{gottwald} with \exos\ and very
recently\cite{wolff} with \rxte, a special type of thermonuclear X-ray
burst that is strong enough to lift up the outer layers of the
star. During these so-called photospheric radius expansion\cite{lewin}
bursts, the radiation flux that emerges from the stellar surface is
limited by the Eddington flux.  It was, however, the
detection\cite{cottam} of redshifted O and Fe lines by \xmm\ from the
surface of the neutron star that rendered this low-mass X-ray binary
unique in its class. For the first time, multiple phenomena have been
observed from a single source that can be concurrently used to
determine uniquely the equation of state of its dense interior. 

The dependence of the three observables, i.e., the Eddington limit,
the redshift, and the ratio $F_{\rm cool}/T_{\rm c}^4$, which is a
quantity closely related to the total emitting area, on the stellar
properties is shown in Table~1. Even though each of these measurements
involve a different combination of the neutron star mass, radius, and
distance, when all three observables are known, the expressions can be
combined and inverted to yield a unique solution for the three stellar
parameters as shown in Table~2. This is also shown visually in
Figure~1, where the different constraints on the neutron star mass and
radius imposed by each observable intersect at a unique set of values.
One additional observable that, in principle, can tighten these
constraints is the amount of rotational broadening in the redshifted
lines. This leads to a direct measurement of the radius of the
star\cite{ozel}, with an uncertainty arising from the angle $i$ that
the observer makes with the rotational axis of the star (Fig. 1).

The only model parameters that are required in this solution are the
color correction factor $f_\infty$ and the electron scattering opacity
$\kappa_{\rm es}$. The color correction factor $f_\infty$, for an
observer at infinity, relates the color temperature to the effective
temperature of the star by $f_\infty \equiv T_{\rm c}/T_{\rm eff}$ and
can be obtained by theoretical modeling of bursting neutron-star
spectra. We use here a fitting function we devised that describes to
an accuracy of 4\% the results of recent model atmosphere  
calculations\cite{madej} for solar-abundance composition 
\begin{equation} 
f_\infty = 1.34 +
0.25\left(\frac{1+X}{1.7}\right)^{2.2} \left(\frac{T^4_{\rm
eff}/10^7{\rm K}}{g/10^{13}{\rm cm}/{\rm s}^2}\right)^{2.2}. 
\end{equation} 
Note that $f_\infty$ in Tables 1 and 2 corresponds to the color
correction factor at the cooling tails of the X-ray bursts when the
emerging flux is substantially sub-Eddington.  At this limit, models
of bursting atmospheres\cite{madej} show that $f_\infty$ asymptotes to
a value of $\sim 1.34$, practically independent of composition,
temperature, and stellar gravity, significantly reducing the model
dependence of our results. For the electron scattering opacity we use
$\kappa_{\rm es} = 0.2 (1+X)$~cm$^2\,$g$^{-1}$, where $0<X<1$ is the
hydrogen mass fraction. This is appropriate for the fully ionized
neutron star atmosphere during the radius-expansion episode. In this
calculation, we allow for the entire range of values for $X$.

The main systematic uncertainty in our results arises from the
assumption that the thermonuclear flash engulfs the entire star
during the radius expansion and cooling phases of the bursts. There
are a number of theoretical and observational arguments that support
this assumption. First, numerical models\cite{spitkovsky} of the
spreading of thermonuclear bursts on the surface of rotating neutron
stars show deflagration times that are $<< 1$~s, which is shorter by
orders of magnitude than the duration of the bursts. Second, the
constraints\cite{loeb} on the magnetic field strength of \exo, imposed
by the lack of Zeeman splitting of the atomic lines, show that the
magnetic field is dynamically unimportant and thus cannot inhibit the
spreading of the thermonuclear flash over the entire surface.

On the observational side, further evidence for uniform emission from
the entire surface as well as the constancy of the Eddington flux
during the bursts is obtained from the study\cite{galloway} of the
peak luminosity of a large number ($\sim 70$) of bursts from
4U1728$-$34, which shows that the peak flux is constant to within
$1\%$ in bursts separated by months. Indeed, a larger
study\cite{gallowayetal} of peak fluxes of all thermonuclear burst
sources in the \rxte\ catalogue also show a quantitatively similar
result.  Second, in the cooling tails of thermonuclear bursts, the
observed ratio $F_{\rm cool}/T_{\rm c}^4$ asymptotes to a constant and
reproducible value between bursts, strongly indicating that the entire
surface of the neutron star emits uniformly in the cooling
tails\cite{strohmayer,damen}.

Even though flux oscillations of amplitude $\lesssim 10\%$ have been
seen in some thermonuclear bursts and are thought to be caused by
modes excited on the neutron star surface during the
bursts\cite{muno}, their presence does not increase the uncertainties
reported here. First, bursts oscillations have never been observed
during the radius expansion phases of the bursts but only in the rise
phase and cooling tails, thus not affecting the determination of the
Eddington limit from observations. Second, the low amplitudes of the
oscillations during the cooling tails introduce uncertainties that are
smaller than both the systematic and statistical uncertainties
($10\%$) allowed for in our calculations. This is especially true for
the burst oscillations in \exo, which are very weak, with an average
amplitude of $3\%$\cite{vs04}.

A final consideration about the Eddington limit is related to the
point during a radius expansion burst at which this quantity is
measured.  The relevant measurement here is the so-called touchdown
flux, which is the Eddington limit at the point in the burst when the
photosphere returns to the actual radius of the star\cite{damen}. The
flux at the touchdown point cannot be smaller than the Eddington flux
because, if the radiation support of the atmosphere were suddenly
removed, the photospheric radius would return to the actual size of
the star within a free-fall timescale, which is $\sim 1$~ms for a
neutron star. Instead, an adiabatic return to the stellar surface, at
timescales of $\sim$~few seconds, is observed in every radius
expansion burst, indicating that the emerging flux traces the
Eddington limit all the way to the touchdown point.

The only unknown property of the binary system that affects the mass
and radius measurements is the hydrogen fraction $X$ of the accreting
material. Further observations of the elemental abundances of the
accretion flow at long wavelengths can greatly reduce this
uncertainty. However, even when we consider the most extreme value of
$X=0.7$, we obtain lower limits on the mass and radius:
\begin{equation} 
M \ge 2.10 \pm 0.28 M_\odot, \qquad R \ge 13.8 \pm 1.8~{\rm km},
\end{equation}
which is shown in Figure~2. For smaller values of the hydrogen mass
fraction, i.e., for a Helium-rich companion which can be expected in a
small-orbit binary such as \exo, the values of $M$ and $R$ are even
higher. It is clear from this figure that the unknown value of $X$ is by 
far the
dominant systematic uncertainty in our result. Also 
note that even though separate uncertainties on mass and radius are 
quoted, these are not independent and do not form an error ellipse 
because the mass-to-radius ratio is fixed by the redshift measurement 
with a negligible error.  

Because we treat distance as an independent variable, we can use the
same three measurements to determine a lower limit on the distance to
the source, $D \ge (9.2 \pm 1.0)$~kpc. This is a significant advantage
of our method: The equations in Table~2 show that if the redshift is
measured, the mass and radius measurements are independent of the
distance and vice versa. Alternatively, if there is a direct
measurement of the distance, e.g., for a source in a globular cluster,
then only two spectroscopic measurements are sufficient to determine
the mass and radius of the neutron star.

The mass-radius measurement presented here does not suffer from
uncertainties related to magnetic and centrifugal support that can
complicate the calculation of neutron star structure given an equation
of state because of the slow rotation and the low magnetic field of
\exo.\ The large size {\it and} the high mass of this neutron star
impose stringent constraints on the equation of state of matter at
supernuclear densities (see Figure~2). Even the lowest allowed value
for the mass and radius can be obtained by only the stiffest equations
of state.  In particular, equations of state with condensate interiors
predict large radius-, low-mass neutron stars (Fig.~2) that are mostly
excluded by our measurements. On the other extreme of the mass-radius
diagram, self-bound bare strange-matter stars have mostly small masses
and radii, also inconsistent with the mass and radius of \exo. The
more conventional equations of state with neutron-proton compositions
are more likely to explain the set of observational properties of
\exo\ presented here. This argues that hadrons and not deconfined 
quarks represent the ground state of matter.

\clearpage

\begin{table}
\begin{center}
\begin{tabular}{>{\small}c >{\small}c >{\small}c}
\hline
Observable & Measurement & Dependence on NS Properties\\
\hline
$\fedd$ & $(2.25 \pm 0.23) \times 10^{-8}$ erg\,\,cm$^{-2}$\,s$^{-1}$ &   
$\frac{1}{4 \pi D^2} \frac{4 \pi GMc}{\kappa_{\rm es}} 
\left(1-\frac{2GM} {c^2R} \right)^{1/2}$  \\
$z$ & 0.35 & $\left(1-\frac{2GM}{Rc^2}\right)^{-1/2} -1 $ \\
$F_{\rm cool}/ \sigma T_{\rm c}^4$ & $1.14 \pm 0.10$ (km/kpc)$^2$ & 
${f_\infty^2} \frac{R^2}{D^2} \left(1-\frac{2GM}{Rc^2}\right)^{-1}$ \\ 
\hline
\end{tabular}
\end{center}

\caption[] {\small The three main quantities observed from \exo\ and
their theoretical dependence on the neutron star properties. The
Eddington limit $F_{\rm Edd}$, defined as the radiation flux at which
the outward radiation force balances the inward gravitational force, is
the limiting flux emerging from thermonuclear X-ray bursts with
photospheric radius expansion. The measurements of the touchdown flux
reported here were obtained by averaging the values determined
recently with \rxte\cite{wolff} and earlier with \exos\cite{gottwald}
observations, which are consistent with each other. The redshift $z$
of O and Fe absorption lines in the X-ray burst spectra of \exo\ has
been measured for the first time with \xmm\cite{cottam}. The ratio
$F_{\rm cool} / \sigma T^4_{\rm c}$, where $F_{\rm cool}$ and $T_{\rm
c}$ are the thermal flux and the color temperature inferred from the
X-ray burst spectra, respectively, asymptotes to a constant value
during the cooling tails of the bursts. This ratio is directly related
to the emitting surface area of the neutron star and was inferred from
the \rxte\ data\cite{wolff} by taking the averages of $F_{\rm cool} /
\sigma T_{\rm c}^4$ measured between 7$-$15~s after the burst,
ensuring that a constant ratio has been reached following the
photospheric radius expansion but that the observed flux is still high
enough for an accurate determination. The physical constants $G$ and
$c$ are the gravitational constant and the speed of light,
respectively, $\kappa_{\rm es} = 0.2 (1+X)$~cm$^2$/g is the electron
scattering opacity, X is the hydrogen mass fraction of the accreted
material, $f_\infty$ is the color correction factor, and $D$ is the
distance to the source. All the above expressions include minimal
general relativistic corrections that are accurate in the slow
rotation limit, which is appropriate for \exo\ given its 47~Hz spin
frequency\cite{vs04}.}  \label{tab:data} \end{table}

\begin{table}
\begin{center}
\begin{tabular}{>{\small}c >{\small}c >{\small}c }
\hline
NS Property & Dependence on Observables & Constraint\\
\hline
M & $\frac{f_\infty^4 c^4}{4 G \kappa_{\rm es}} 
\left(\frac{F_{\rm cool}}{\sigma 
T_{\rm c}^4}\right) \frac{[1-(1+z)^{-2}]^2}{(1+z)^3} F_{\rm Edd}^{-1} $ & 
$2.10 \pm 0.28 M_\odot $ \\
R & $\frac{f_\infty^4 c^2}{2 \kappa_{\rm es}} 
\left(\frac{F_{\rm cool}}{\sigma 
T_{\rm c}^4}\right) \frac{1-(1+z)^{-2}}{(1+z)^3} F_{\rm Edd}^{-1} $  & 
13.8$\pm$1.8 km  \\
D & $\frac{f_\infty^2 c^2}{2 \kappa_{\rm es}} 
\left(\frac{F_{\rm cool}}{\sigma 
T_{\rm c}^4}\right)^{1/2} \frac{1-(1+z)^{-2}}{(1+z)^3} F_{\rm Edd}^{-1} $ & 
9.2$\pm$1.0 kpc \\
\hline
\end{tabular}
\end{center}
\caption[]{\small The neutron star properties that are obtained from
the observations summarized in Table~1. The neutron star mass, radius,
and distance are uniquely determined from the Eddington luminosity,
the redshift, and the ratio $F_{\rm cool}/ \sigma T_{\rm c}^4$ as
shown in the middle column, given a model for the neutron star
atmosphere that determines the color correction factor and a
measurement of the hydrogen mass fraction. The last column shows the
minimum values for the mass, radius, and distance to the neutron star
for any possible value of the color correction factor and the hydrogen
mass fraction. The counter-intuitive dependence of the stellar
properties on the observables arises from the combination of the
expressions shown in Table~1. These expressions also show why a unique
spectroscopic determination of the mass and radius of a neutron star,
which is necessary to constrain its equation of state, can only be
achieved when all three phenomena are observed from a single source, if 
the distance to the source is unknown.}
\label{tab:constraints}
\end{table}

\clearpage


\begin{acknowledge} 

I am grateful to the gamma ray group, especially Chryssa Kouveliotou,
for their hospitality at MSFC where this idea was born; and for the
hospitality of the members of Anton Pannekoek Institute at the
University of Amsterdam, where the work was completed. I thank
Dimitrios Psaltis for useful discussions and comments on the
manuscript.

\end{acknowledge}

\clearpage

\begin{figure} 
\centerline{\psfig{file=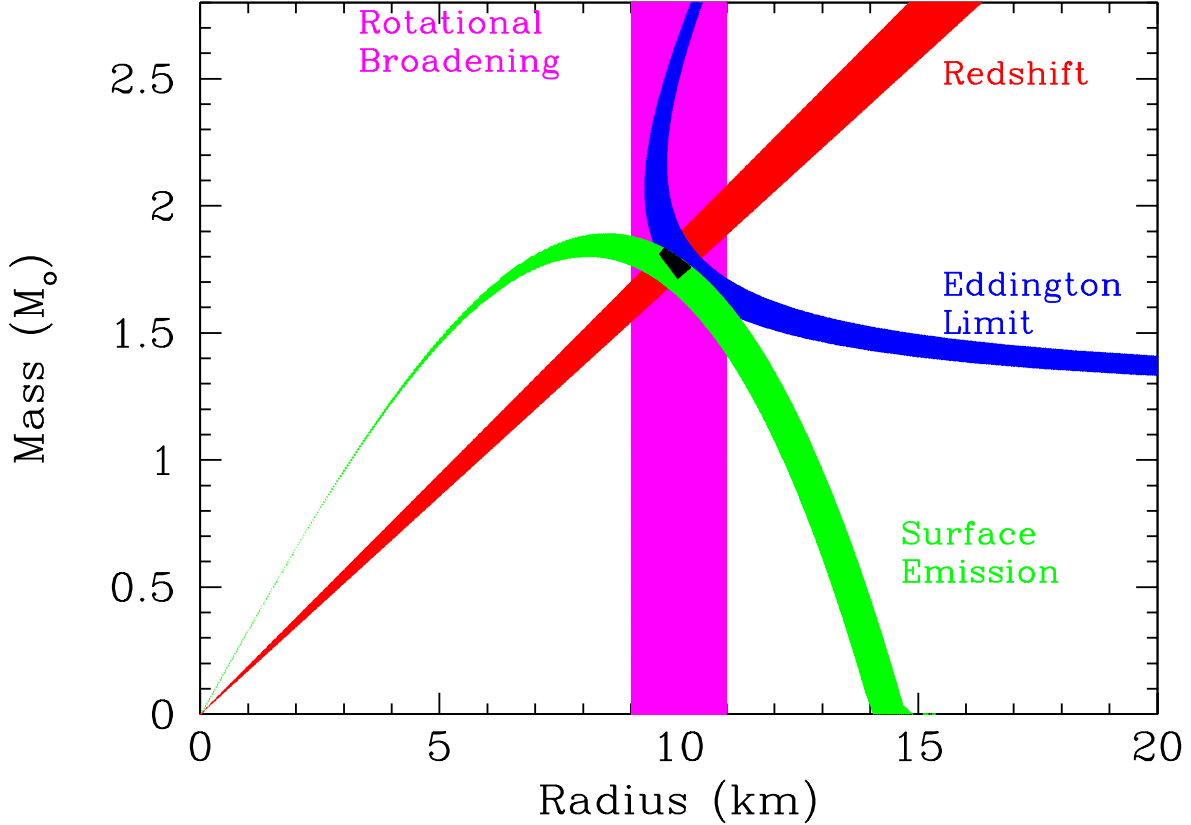,width=6.5in,angle=-90}}
\caption[]{\small {\bf Four complementary methods to
determine the mass and radius of a neutron star.} The contours on the
mass-radius plane of neutron stars imposed by the measurement of {\it
(i)} the Eddington flux during photospheric radius expansion bursts
(blue), {\it (ii)} the ratio $F_{\rm cool}/T_{\rm c}^4$ of the surface
emission that asymptotes to a constant during the cooling tail of a
thermonuclear burst (green), {\it (iii)} the redshift of atomic
absorption lines observed during the burst (red), and {\it (iv)} the
broadening of such lines due to the rotation of the star (magenta) for
a hypothetical star with $M=1.8 M_\odot$ and $R=10$~km.  The second
quantity, obtained from the thermal flux $F_{\rm cool}$ and the color
temperature $T_{\rm c}$ of the burst spectrum, is closely related to
the total emitting area from the stellar surface when the nuclear
burst has engulfed the entire star.  The widths of the contours
correspond to a hypothetical $10\%$ uncertainty in each
measurement. The uncertainty in the redshift measurement can be
negligible if a grating spectrometer is used, as in the case of \exo.\
The black shaded area is the intersection of the four contours and
corresponds to the uncertainty in the measurement of the true mass and
radius of the neutron star. } \label{fig:fig1} 
\end{figure}

\clearpage

\begin{figure} 
\centerline{\psfig{file=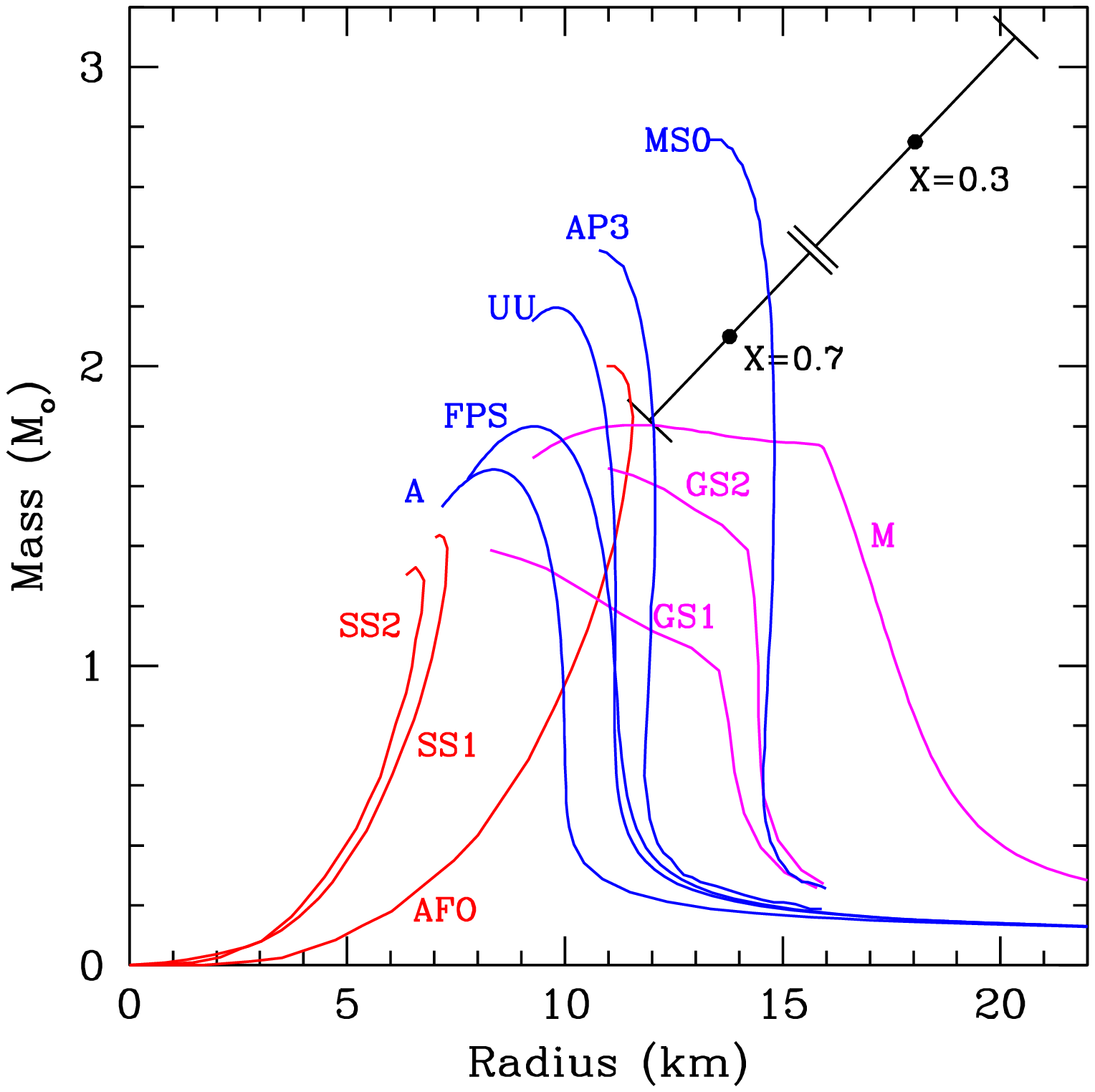,width=5.6in,angle=0}}
\caption[]{\small {\bf The constraints on the neutron
star equations of state imposed by observations of \exo.} The
predicted mass-radius relations for a number of representative
equations of state of neutron stars without condensates (blue), with
condensates (magenta), and for strange stars
(red)\cite{lattimer,cook}. The curve labels and the corresponding
references can be found in refs.~[15] and [16].  The two $1-\sigma$
error bars corresponds to the measurement of the mass and radius of
\exo\ from three independent observations (equation~2) for different
values of the hydrogen mass fraction $X$ of the accreted material. The
measurement for $X=0.7$ corresponds to the minimum allowed values for
the mass and radius of the neutron star. Only the stiffest equations
of state are consistent with this measurement. } \label{fig:fig2}
\end{figure}


\begin{thebibliography}{10}

\bibitem[{hands}<1>]{hands}
{Hands}, S. 
{The Phase Diagram of QCD.}
\newblock Contemp. Phys. {\bf 42}, 209 (2001)

\bibitem[{Lattimer} \& {Prakash}<2>]{lp} 
{Lattimer}, J.~M. \& {Prakash}, M.
{The Physics of Neutron Stars.}
\newblock {\it Science} {\bf 304}, 536 (2004)

\bibitem[{Gottwald} {\it et~al}.<3>]{gottwald} 
{Gottwald}, M., {Haberl}, F., {Parmar}, A.~N., \& {White}, N.~E.
{The bursting behavior of the transient X-ray burst source 
\exo\ - A dependence between the X-ray burst properties 
and the strength of the persistent emission.}
\newblock {\it Astrophys. J.} {\bf 308}, 213-224 (1986)

\bibitem[{Wolff} {\it et~al.}<4>]{wolff}
{Wolff}, M.~T., {Becker}, P.~A., {Ray}, P.~S., \& {Wood}, K.~S.
{A Strong X-ray burst from the low mass X-ray binary EXO 0748-676}.
\newblock {\it Astrophys. J.}, in press

\bibitem[{Lewin} {\it et~al}<5>]{lewin}
{Lewin}, W.~H.~G., {van Paradijs}, J., \& {Taam}, R. 
{X-ray Bursts}
\newblock {X-ray Binaries, eds. W.~H.~G.\ Lewin, E.~P.~J.\ van den 
Heuvel, \& J.\ van Paradijs, (University Press: Cambridge)} (1995)

\bibitem[{Cottam} {\it et al}.<6>]{cottam} 
{Cottam}, J., {Paerels}, F., \& {Mendez}, M. 
{Gravitationally redshifted absorption lines in the X-ray burst 
spectra of a neutron star}.
\newblock {\it Nature} {\bf 420}, 51-54 (2002)

\bibitem[{\"O}zel and Psaltis <7>]{ozel} 
{\"O}zel, F., {Psaltis}, D.\ 
{Spectral Lines from Rotating Neutron Stars}.
\newblock {\it Astrophys. J.} {\bf 582}, L31-L34 (2003)

\bibitem[{Madej} {\it et~al.}<8>]{madej} 
{Madej}, J., {Joss}, P.~C., {R{\'o}{\.z}a{\'n}ska}, A. 
{Model Atmospheres and X-Ray Spectra of Bursting Neutron Stars: 
Hydrogen-Helium Comptonized Spectra}.
\newblock {\it Astrophys. J} {\bf 602}, 904-912 (2004)

\bibitem[{Spitkovsky} {\it et~al}.<9>]{spitkovsky} 
{Spitkovsky}, A., {Levin}, Y., \& {Ushomirsky}, G.
{Propagation of Thermonuclear Flames on Rapidly Rotating Neutron 
Stars: Extreme Weather during Type I X-Ray Bursts.}
\newblock {\it Astrophys. J.} {\bf 566}, 1018-1038 (2002)

\bibitem[{Loeb}<10>]{loeb} 
{Loeb}, A.
{Spectroscopic Constraints on the Surface Magnetic Field of the
Accreting Neutron Star EXO0748-676}.
\newblock {\it Physical Review Letters} {\bf 91}, 071103 (2004)

\bibitem[{Galloway} {\it et~al.}<11>]{galloway} 
{Galloway}, D.~K., {Psaltis}, D., {Chakrabarty}, D., {Muno}, M.~P.\ 
{Eddington-limited X-Ray Bursts as Distance Indicators. I. Systematic 
Trends and Spherical Symmetry in Bursts from 4U~1728-34}.
\newblock {\it Astrophys. J.} {\bf 590}, 999-1007 (2003)

\bibitem[{Galloway et al}<12>]{gallowayetal} 
{Galloway}, D.~K., {Muno}, M.~P., {Chakrabarty}, D., {Psaltis}, D., {Hartman}, J.~M.\  
{Thermonuclear Bursts Observed by RXTE: the MIT catalogue}.
\newblock {\it HEAD} {\bf 8.2520G} (2004)

\bibitem[{strohmayer} {et~al.}<13>]{strohmayer}
{Strohmayer}, T.~E., {Zhang}, W., \& {Swank}, J.~H.\ 
{363 Hz oscillations during the rising phase of bursts from 4U~1728-34:
Evidence for rotational modulation}
\newblock {\it Astrophys. J} {\bf 487}, L77-80 (1997)

\bibitem[{Damen} {\it et~al}.<14>]{damen} 
{Damen}, E., {Magnier}, E., {Lewin}, W.~H.~G., {Tan}, J., {Penninx}, W., 
{van Paradijs}, J.\  
{X-ray bursts with photospheric radius expansion and the gravitational 
redshift of neutron stars}.
\newblock {\it Astron. and Astrophys.} {\bf 237}, 103-109 (1990)

\bibitem[{Muno} {\it et~al}.<15>]{muno} 
{Muno}, M.~P., {\"O}zel, F., \& {Chakrabarty}, D.
{The Amplitude Evolution and Harmonic Content of Millisecond 
Oscillations in Thermonuclear X-Ray Bursts}.
\newblock {\it Astrophys. J.} {\bf 581}, 550-561 (2002)

\bibitem[{Villarreal} \& {Strohmayer}<16>]{vs04}
{Villarreal}, A.~R. \& {Strohmayer}, T.~E. 
{Discovery of the Neutron Star Spin Frequency in EXO 0748-676}.
\newblock {\it Astrophys. J.} {\bf 614}, L121-L124 (2004)

\bibitem[{Lattimer} \& {Prakash}<17>]{lattimer} 
{Lattimer}, J.~M. \& {Prakash}, M.
{Neutron Star Structure and the Equation of State.}
\newblock {\it Astrophys. J.} {\bf 550}, 426-442 (2001)

\bibitem[{Cook} et al.(1994)<18>]{cook} 
Cook, G.~B., Shapiro,S.~L., Teukolsky, S.~A.\ 
{Rapidly rotating neutron stars in general relativity: 
Realistic equations of state.}\ 
\newblock {\it Astrophys. J.} {\bf 424}, 823-845 (1994)

\end{thebibliography}
\end{document}